# The Effects of Problem Posing Learning Model on Students' Learning Achievement and Motivation

**Agatha Puri Christidamayani[1], Yosep Dwi Kristanto[2]**
[1,2] Faculty of Teacher Training and Education, Universitas Sanata Dharma, Indonesia

**Abstract**

Posing high-quality problems is a critical skill to be possessed by students in learning mathematics. However, it is still limited in literature in answering whether problem posing learning model effective in improving students' learning achievement and motivation. Therefore, the present study aims to investigate the effect of problem posing learning model in the topics of cube and cuboid on students' learning achievement and motivation. This study employs quasi experimental design with 20 students in experimental group and 24 students in control group. The study found that the problem posing learning model has insignificant effect on the students' learning achievement but has a positive and significant effect of the learning model on the students' learning motivation. Further analysis showed that the learning model also has a significant and positive effect on every aspect of students learning motivation, namely students' interests, enthusiasm, diligence, collaboration, and self-control.

*Corresponding Author:*
*Yosep Dwi Kristanto, Faculty of Teacher Training and Education, Universitas Sanata Dharma, Indonesia*
*Email: yosepdwikristanto@usd.ac.id*

## 1. Introduction

Posing high-quality problems is critical skill to be possessed by students in learning mathematics. Engaging students in formulating problems is considered as an effective effort to improve students learning (Cai & Jiang, 2017). Furthermore, formulating problems is an important step in scientific investigation. As Einstein stated, "the formulation of a problem is often more essential than its solution" (Einstein & Infeld, 1938). Moreover, Socrates has shared a learning method in which the students actively engage in posing problems critically (Singer, Ellerton, & Cai, 2013). Recognizing the importance of the ability to pose problems for students, recently researchers give an emphasize that students need to have more active role in their learning by means of posing problems during the problem posing activities (Kalmpourtzis, 2019; Putra, Herman, & Sumarmo, 2020; Ye, Chang, & Lai, 2019).

Various definitions of problem posing are found in the literature. Based on widely cited definition by Silver (1994), problem posing includes the process of new problems generation and process of re-formulating existing problem. In more detail, Hobri (2008) define problem posing as (a) simple formulation of questions or re-formulation of existing problems with some changes so that they are simpler and can be mastered; (b) formulation of questions relating to the conditions of the questions that have been solved to find alternative solutions; (c) formulation of the problem from the information or situation available, whether done before, when, or after solving the problem. Other definitions of problem posing have appeared in Cai and Hwang (2019).

Problem posing learning model provides benefits for students. Upu (2003) explained that problems posing is one approach that can increase active student involvement in the mathematics learning process. Proposing a problem can be useful in bringing together a number of learning goals, both in group and individual learning setting. Furthermore, Cai, Hwang, Jiang, & Silber (2015) posit that problem-posing activities can improve students' learning achievement, reasoning and communication skills, and capture their motivation.

To sum up, literature give an insight that the problem posing learning model has a potential in improving students' learning achievement and motivation, even though it still limited. Therefore, the research questions of the present study are as follows: (1) Does problem-posing learning model have an effect on the students' learning achievement? (2) Does problem-posing learning model have an effect on the students' learning motivation?

In problem posing setting, students need to generate new problems or re-formulate the existing problems. To this end, students require to reflect on their prior knowledge and understanding and connect them by using problem posing activities (Moses, Bjork, & Goldenberg, 1990). This cognitively demanding process results in a deeper understanding on mathematical concepts. Therefore, our first hypothesis is:

> *Hypothesis 1*: Problem-posing learning model have a positive effect on the students' learning achievement.

Problem posing learning model facilitates students to have an active role in the learning process. This active involvement in their own learning will make the students more confidence and have positive attitudes (Guvercin, Cilavdaroglu, & Savas, 2014), and in turn, their motivation levels increase. Hence, our second hypothesis is:

> *Hypothesis 2*: Problem-posing learning model have a positive effect on the students' learning motivation.

## 2. Method
### a. Research Design

The present study employed pre-test post-test control group design, which is one type of quasi-experimental designs, in investigating the effects of the independent variable (i.e. with and without problem-posing model) on dependent variables (i.e. students' learning achievement and motivation). With this research design, the causal relationship between independent and dependent variables can be determined since the data are observed under the control of researcher. This study consisted of three meetings in mathematics lesson in the topic of cube and cuboid.

### b. Subjects

The subjects in the present study were 44 eight-grade students of St. Vincentius Pangudi Luhur Middle School, Yogyakarta. They were randomly assigned to the experimental and control group. As a result, the experimental group included 20 students whereas the control group consisted 24 students.

### c. Data Collection Tools

In determining the effects of problem posing learning model on students' learning achievement and motivation, first, we ensured the implementation of problem posing model in the experimental group and vice versa in the control group by using observation sheets.

The observation sheets were developed based on the lesson plans for each group. The lesson plans in problem posing (experimental) group have five key phases, namely informing learning goals, groups formation, presenting problems, posing problems, and solving the problems. The observation on learning implementation was conducted by 2 independent observers.

The learning achievement test was used to measure the students' learning achievement (see, Christidamayani, 2019). This test which consisted five items at the levels of knowledge, understanding, and application was validated by two experts. The revision was made based on the comments from the experts.

In measuring students' learning motivation, the learning motivation questionnaire was developed. The questionnaire consisted five aspects, namely students' interest, enthusiasm, diligence in reviewing materials, identity, as well as collaboration and self-control. These aspects were adopted from Brown (as cited by Imron, 1996). The questionnaire has been validated by two experts and revised based on their comments. The Cronbach's Alpha reliability coefficient of the questionnaire was 0.881.

### d. Data Analysis

The data of the learning implementation were presented as proportion for each learning phase. The data of students' learning achievement and motivation were tested by the Kolmogorov-Smirnov test in determining their normality. The statistical analyses conducted in this study were independent sample *t*-test and Mann-Whitney U test. The data analyses were conducted through SPSS Statistics 23 and Minitab 19.

The normality test by the Kolmogorov-Smirnov test on students' learning achievement pre-test, students' motivation pre- and post-questionnaire showed that the data were normally distributed. However, the same test was conducted on students' learning achievement post-test and resulted that the data did not normally distributed.

The normality test also has been conducted on five aspects of students' learning motivation before and after the implementation. These resulted that all scores were normally distributed except aspect 1 before the implementation for both experimental and control groups, aspect 2 before and after for experimental groups, and aspect 3 after the implementation for experimental group.

### 3. Result and Discussion

The results of data analyses on learning implementation, students' learning achievement and motivation are described in the following subsections.

### a. Results Regarding the Learning Implementation

Results of learning implementation in both experimental and control groups are showed in Table 1. Based on the Table 1, it is found that the learning implementation proportion for each phase is no less than 83%. These results are prerequisites for the further analysis on students' learning achievement and motivation. Since the proportion of the learning implementation is high, the next analyses can be conducted.

Table 1. Proportion of Learning Implementation in Each Phase

|  | Proportion of Learning Implementation | | |
|---|---|---|---|
|  | Opening | Core | Closure |
| Control Group |  |  |  |
| Meeting I | 100% | 94% | 83% |
| Meeting II | 100% | 100% | 100% |

|  | Proportion of Learning Implementation | | |
|---|---|---|---|
|  | Opening | Core | Closure |
| Meeting III | 100% | 100% | 100% |
| Experimental Group |  |  |  |
| Meeting I | 100% | 83% | 88% |
| Meeting II | 100% | 100% | 100% |
| Meeting III | 93% | 100% | 100% |

**b. Results on Students' Learning Achievement**

Results of two-tailed independent sample *t*-test (equal variances not assumed) on students' learning achievement pre-test scores were given in the Table 2.

According to Table 2, there were no significant difference between students' learning achievement pre-tests scores in experimental and control group. Therefore, it comes to conclusion that the students' learning achievement in experimental and control groups were equal at the beginning of the present study.

Table 2. Results of t-Test on Students' Learning Achievement Pre-test Scores

| Group/Test | N | M | SD | t | p |
|---|---|---|---|---|---|
| Experimental Group Pre-test | 20 | 54.20 | 14.48 | 0.674 | 0.504 |
| Control Group Pre-test | 24 | 57.42 | 17.17 |  |  |

Table 3 showed the results of one-tailed Mann-Whitney U test on students' achievement post-tests scores. Based on the Table 3, it can be said that students' learning achievement in experimental group is not significantly higher than in control group.

Therefore, although the problem posing learning model seems to have a positive effect on the student learning achievement, but the effect is not significant. This result does not support hypothesis 1.

Table 3. Results of Mann-Whitney U Test on Students' Learning Achievement Post-test Scores

| Group/Test | N | M | SD | U | p |
|---|---|---|---|---|---|
| Experimental Group Pre-test | 20 | 84.10 | 11.65 | 531.00 | 0.420 |
| Control Group Pre-test | 24 | 79.96 | 17.78 |  |  |

**c. Results of Students' Learning Motivation**

The two-tailed independent samples *t*-test of students' learning motivation scale before the implementation and the one-tailed independent samples *t*-test of the corresponding scale after the implementation were given in the Table 4. Based on Table 4, there were no significant difference between the students' learning motivation in experimental and control groups before the implementation. Therefore, the students came with equal motivation at the beginning of the study. Furthermore, it can be concluded from Table 4 that students' learning motivation in experimental group was significantly higher than the students in control group after the implementation. Therefore, the problem posing model has a significant positive effect on students' learning motivation. This result support hypothesis 2.

Table 4. Results of t-Test on Students' Learning Motivation Before and After the Implementation

| Group/Test | N | M | SD | t | p |
|---|---|---|---|---|---|
| Before the Implementation | | | | | |
|   Experimental Group | 20 | 83.25 | 19.90 | 1.427 | 0.162[a] |
|   Control Group | 24 | 91.33 | 17.16 | | |
| After the Implementation | | | | | |
|   Experimental Group | 20 | 104.40 | 13.53 | 2.980 | 0.003[b] |
|   Control Group | 24 | 89.25 | 20.02 | | |

To explore the effect of problem posing model on students' learning motivation in more detail, Table 5 shows the results of two-tailed $t$-test and Mann-Whitney U test on students' learning motivation scale before the implementation.

Table 5. Results of t-Test and Mann-Whitney U Test for Each Aspect of Students' Learning Motivation Before the Implementation

| Aspect/Group | M | SD | t | U | p |
|---|---|---|---|---|---|
| Aspect 1 | | | | | |
|   Experimental Group | 18.70 | 4.46 | 0.586 | | 0.561 |
|   Control Group | 17.96 | 3.82 | | | |
| Aspect 2 | | | | | |
|   Experimental Group | 15.85 | 3.91 | | 632.50 | 0.030 |
|   Control Group | 19.08 | 5.09 | | | |
| Aspect 3 | | | | | |
|   Experimental Group | 14.55 | 5.63 | −1.027 | | 0.311 |
|   Control Group | 16.21 | 4.95 | | | |
| Aspect 4 | | | | | |
|   Experimental Group | 14.40 | 6.18 | −1.211 | | 0.233 |
|   Control Group | 16.54 | 5.40 | | | |
| Aspect 5 | | | | | |
|   Experimental Group | 19.75 | 3.78 | −1.609 | | 0.116 |
|   Control Group | 21.54 | 3.55 | | | |

According to the Table 5, there were no significant difference between students' learning motivation scale before the implementation in experimental and control groups for aspect 1, 3, 4, and 5. However, there were significant difference for aspect 2.

Therefore, students had equal interest, diligence, identity, collaboration, and self-control at the beginning of the study, but with different enthusiasm.

Table 6 gives the results of one-tailed independent samples $t$-test and Mann-Whitney U test of students' learning motivation scale after the implementation for aspect 1, 3, 4, and 5. Based on the Table 6, it can be concluded that the students' motivation on aspect 1, 3, 4, and 5 in experimental group were significantly higher than the students' in control group. Thus, the problem posing model has a significant and positive effect on students' interest, diligence, identity, collaboration, and self-control.

Table 6. Results of t-Test and Mann-Whitney U Test for Each Aspect of Students' Learning Motivation After the Implementation

| Aspect/Group | M | SD | t | U | p |
|---|---|---|---|---|---|
| Aspect 1 | | | | | |
|    Experimental Group | 20.60 | 2.70 | | 435.50 | 0.007 |
|    Control Group | 18.29 | 3.86 | | | |
| Aspect 2 | | | | | |
|    Experimental Group | 20.00 | 3.81 | | 459.50 | 0.029 |
|    Control Group | 17.08 | 5.57 | | | |
| Aspect 3 | | | | | |
|    Experimental Group | 19.55 | 4.29 | | 445.50 | 0.013 |
|    Control Group | 15.67 | 5.51 | | | |
| Aspect 4 | | | | | |
|    Experimental Group | 20.10 | 4.27 | 2.217 | | 0.016 |
|    Control Group | 17.42 | 4.27 | | | |
| Aspect 5 | | | | | |
|    Experimental Group | 24.15 | 2.54 | 2.518 | | 0.009 |
|    Control Group | 20.79 | 5.91 | | | |

Further analysis conducted on aspect 2 of students' motivation. By utilizing one tailed Mann-Whitney U test on before-implementation scale, it was resulted U = 632.5 and $p$ = 0.015. Therefore, the students' enthusiasm before the implementation in control group were significantly higher than students' in experimental group. The one-tailed Mann-Whitney U test was also conducted on the corresponding after-implementation scale, resulting in U = 459.5 and $p$ = 0.029. Thus, the students' enthusiasm in experimental group were significantly higher than students' in control group after the implementation. The illustration of these results is showed in Figure 1.

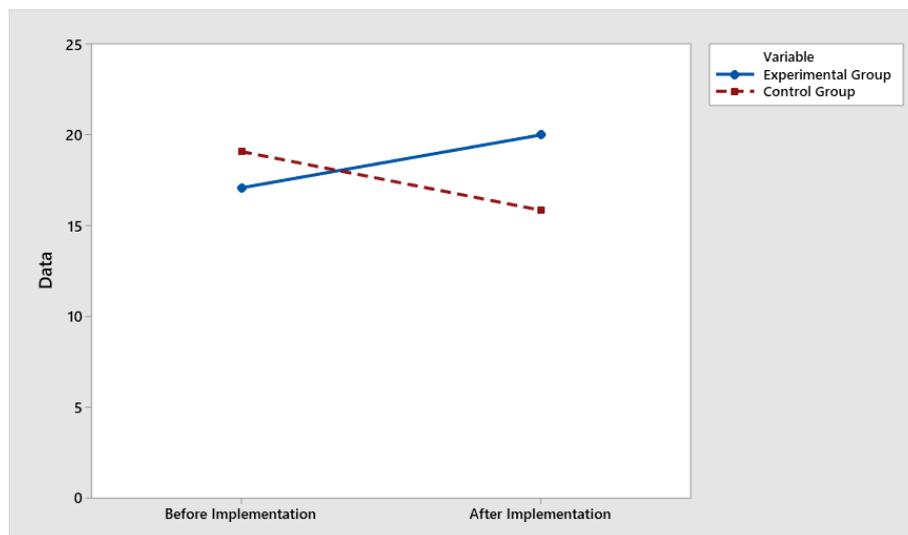

Figure 1. Means Graph of Students' Enthusiasm

Figure 1 shows a cross-over pattern. That is, students' enthusiasm in experimental group starting out significantly lower than the control group and ending up above them. This is the evidence that the problem posing learning approach effective in improving students' enthusiasm.

In the present study we investigate the effects of problem posing learning model on students' learning achievement and motivation. The results indicate that the problem posing learning model does not have any statistically significant positive effect on students' learning achievement, but it does on students' motivation. The students in problem posing group had higher motivation score than their non-problem posing peers.

One possible explanation on why the problem posing in the present study does not have significant positive effect on students' achievement is the lack of students' experience on problem posing learning model. It is also one of main challenges in implementing problem posing learning model (Hsiao, Hung, Lan, & Jeng, 2013). Furthermore, Yu, Liu, and Chan (2005) added that the posing problems were difficult task for the low-achiever students. With regard to these challenges, support is needed for students in problem posing learning environment.

Our finding on the positive effect of problem posing on students' motivation is in line with other studies (Irvine, 2017). On top of the problem posing's positive effect on students' motivation, it also has a same effect on all aspects of motivation, namely interest, enthusiasm, diligence, identity, collaboration, and self-control. Based on this finding and the importance of motivation on students' learning (Linnen-brink, 2007; Tella, 2007; Wijayanti, 2019), the problem posing learning model is a promising strategy to facilitate students' learning.

Finally, the present study findings give an insight for teachers or instructors in implementing problem posing learning model. The teachers who interested in implementing problem posing learning model should pay attention to learning components that affect its effectiveness. As mentioned before, support should be provided to the students, especially for inexperienced and low-achiever students. Teachers also may structure the problem posing learning model with innovative teaching strategies and technologies, e.g. worked example (Hsiao, Hung, Lan, & Jeng, 2013), worksheet scaffold (Choo, Rotgans, Yew, & Schmidt, 2011), game-based problem-solving (Chang, Wu, Weng, & Sung, 2012), "what if not?" strategy (Brown & Walter, 1983; Lavy & Bershadsky, 2003), and Desmos mathematically rich activities (Kristanto, 2019).

4. **Conclusion**

The present study gives a contribution in term of the problem posing's effect on students' learning. The evidence of its positive effect on students' motivation in learning mathematics gives insight for mathematics teachers and educators in improving the effectiveness of mathematics teaching and learning. This study also shows the need for support for inexperienced and low-achiever students in the implementation of problem posing learning model. Moreover, structuring problem posing process by adopting innovative strategies and technologies may make problem posing more effective in facilitating students' learning.